# Giant resonant enhancement for photo-induced superconductivity in $K_3C_{60}$

E. Rowe[1], B. Yuan[1], M. Buzzi[1], G. Jotzu[1], Y. Zhu[1], M. Fechner[1], M. Först[1], B. Liu[1,2]

D. Pontiroli[3], M. Riccò[3], A. Cavalleri[1,4]

[1] *Max Planck Institute for the Structure and Dynamics of Matter, Hamburg, Germany*
[2] *Paul Scherrer Institute, Villigen, Switzerland*
[3] *Dipartimento di Scienze Matematiche, Fisiche e Informatiche, Università degli Studi di Parma, Italy*
[4] *Department of Physics, Clarendon Laboratory, University of Oxford, United Kingdom*

**Photo-excitation at terahertz and mid-infrared frequencies has emerged as a new way to manipulate functionalities in quantum materials, in some cases creating non-equilibrium phases that have no equilibrium analogue. In $K_3C_{60}$, a metastable zero-resistance phase was documented with optical properties and pressure dependences compatible with non-equilibrium high temperature superconductivity. Here, we report the discovery of a dominant energy scale for this phenomenon, along with the demonstration of a giant increase in photo-susceptibility near 10 THz excitation frequency. At these drive frequencies a metastable superconducting-like phase is observed up to room temperature for fluences as low as ~400 µJ/cm². These findings shed light on the microscopic mechanism underlying photo-induced superconductivity. They also trace a path towards steady state operation, currently limited by the availability of a suitable high-repetition rate optical source at these frequencies**

The search for new non-equilibrium functional phases in quantum materials, such as optically induced ferroelectricity[1,2], magnetism[3-5], charge density wave order[6,7], non-trivial topology[8,9] and superconductivity[10-18], has become a central research theme in condensed matter physics. In the case of $K_3C_{60}$ (Fig. 1a), mid infrared optical pulses have been extensively documented to yield an unconventional non-equilibrium phase which exhibits metastable zero-resistance[14], an extraordinarily high mobility and a superconducting-like gap in the optical conductivity[12,14] that reduce with applied pressure[13], and nonlinear I-V characteristics[19]. All these observations are indicative of non-equilibrium high temperature superconductivity, observed at base temperatures far exceeding the highest equilibrium superconducting critical temperature of any alkali-doped fulleride (Fig. 1b).

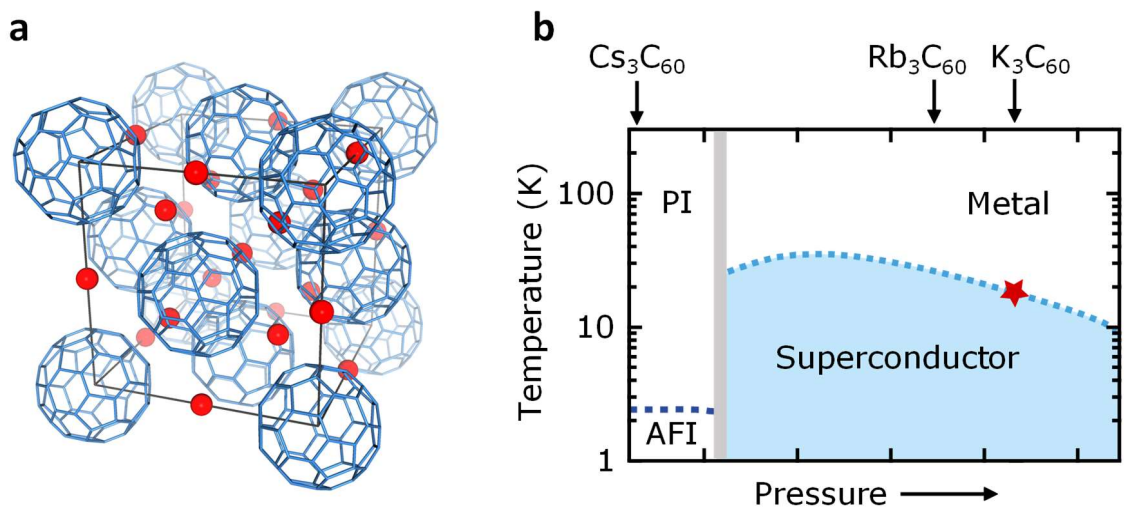

**Figure 1. Crystal structure and phase diagram $K_3C_{60}$.** **(a)** Crystal structure of the organic molecular solid $K_3C_{60}$. $C_{60}$ molecules are situated at the vertices of a face-centered-cubic lattice. Potassium atoms (red) occupy the interstitial voids. **(b)** Pressure-temperature phase diagram of the fcc-$A_3C_{60}$ alkali-doped fulleride family of compounds. Physical pressure tunes the spacing between the $C_{60}$ molecules. The grey line indicates the boundary between the insulating and metallic/superconducting compounds. The blue shaded area indicates where superconductivity is observed at equilibrium. The star indicates the $K_3C_{60}$ compound investigated in this work, which superconducts at temperatures $T \leq T_c = 20K$.

Typical experimental results reported to date are displayed in Fig. 2c. $K_3C_{60}$ powders were held at a base temperature $T = 100$ K $\gg T_c = 20$ K and irradiated with 1 ps-long pulses with 170 meV photon energy ($\lambda \approx 7.3$ µm, $f \approx 41$ THz) at a fluence of 18 mJ/cm². This strong excitation regime yielded a long-lived transient state with dramatic changes in *both* the real and imaginary parts of the optical conductivity, measured using phase sensitive terahertz time-domain spectroscopy.

The transient optical properties displayed in Fig. 2c are reminiscent of those of the equilibrium superconducting state measured in the same material at $T \ll T_c = 20$ K (cf. Fig. 2b), and are suggestive of transient high temperature superconductivity. These signatures consist of perfect reflectivity, a gap in the real part of the optical conductivity $\sigma_1(\omega)$, and an imaginary conductivity $\sigma_2(\omega)$ which diverges towards low frequencies as $\sigma_2(\omega) \propto 1/\omega$. The divergent $\sigma_2(\omega)$ implies (through Kramers-Kronig relations) the presence of a peak in $\sigma_1$ centred at zero frequency, with a width limited by the lifetime of the state which also determines the carrier mobility.

This data was obtained by accounting for the inhomogeneous excitation of the probed volume using a multilayer model. Here we show the results of this reconstruction under the assumption of a linear (filled blue circles) and sublinear (open symbols)[20] dependence of the photo-induced changes in the terahertz refractive index on the mid-infrared pump fluence. Two sub-linear models are shown, with an assumed square-root fluence dependence in blue circles and saturating fluence dependence in brown triangles, as detailed in supplementary section S6. Allowing for a finite temperature superconductor, in which a varying density of uncondensed quasi-particles also contributes to the terahertz response, the superconducting-like nature of the transient state is independent of the specific choice of assumption[21]. Only quantitative differences, associated with the relative densities of induced superfluid and heated quasi-particles, which can be extracted by fitting with a two-fluid model, emerge. This modelling also reveals that the enhancement of conductivity observed in these experiments is not connected to an increase in the carrier density, which remains constant upon photo-excitation, but rather a colossal increase in the carrier mobility.

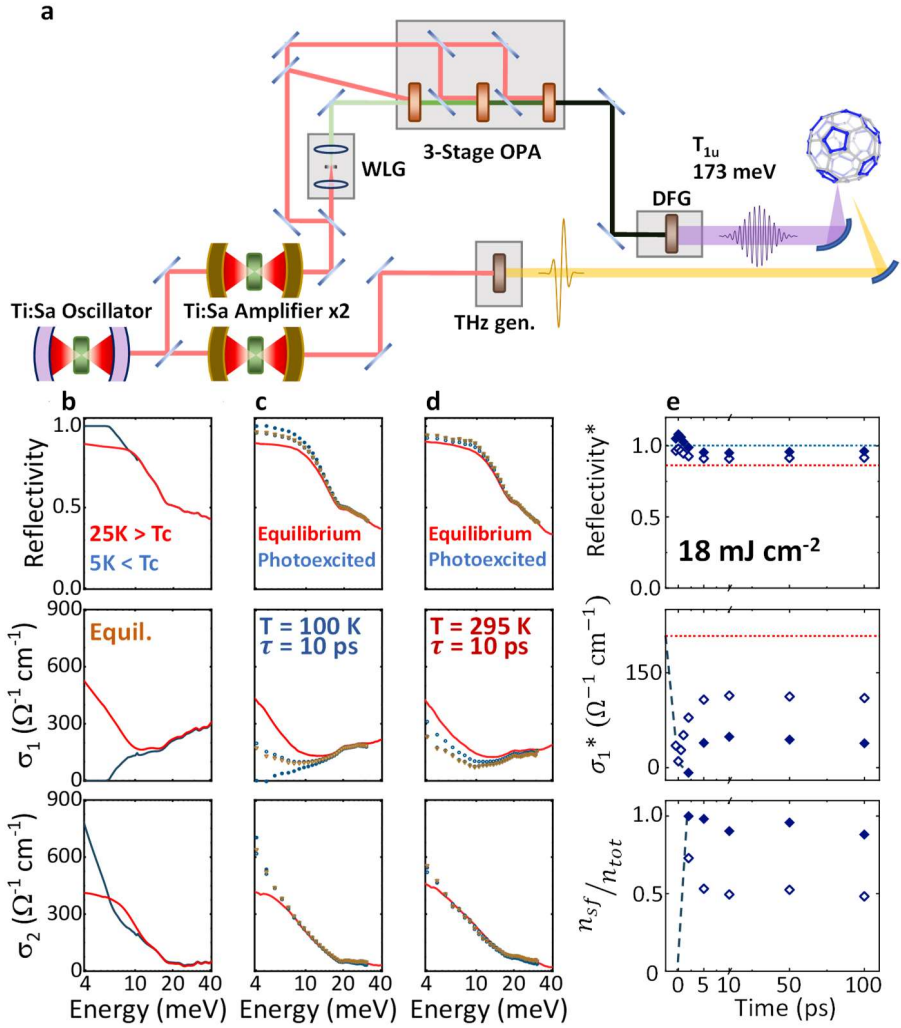

**Figure 2: Photo-induced metastable superconductivity in $K_3C_{60}$ generated with intense 170 meV excitation pulses. (a)** Schematic of the experimental set-up. Pump pulses with 170 meV (41 THz) photon energy (close to resonance with the illustrated phonon mode) were generated via optical parametric amplification (OPA) and subsequent difference frequency generation (DFG) of the signal and idler beams. These pulses were stretched to a duration of $\Delta\tau_{pump} \approx 1$ ps by linear propagation in a highly dispersive $CaF_2$ rod. The photoinduced changes in the far-infrared optical properties of $K_3C_{60}$ were detected with phase-sensitive transient THz time-domain spectroscopy. **(b)** Reflectivity (sample–diamond interface), real and imaginary part of the optical conductivity of $K_3C_{60}$ measured upon cooling across the equilibrium superconducting transition. **(c)** Same quantities measured at equilibrium (red lines) and 10 ps after excitation (symbols). The data in filled blue circles, open blue circles and open brown triangles was obtained accounting for the inhomogeneous excitation of the probed volume under the assumption of a linear, square root and saturating (respectively) fluence dependence of the photo-induced changes in the complex refractive index of the material (Supplementary Section S6). This data was acquired at a base temperature T = 100 K with an excitation fluence of approx. 18 mJ cm$^{-2}$. **(d)** Same quantities as in (c) but measured at a base temperature T = 295 K with an excitation fluence of approx. 15 mJ cm$^{-2}$. **(e)** Time dependence of the average value of reflectivity and real part of the optical conductivity $\sigma_1(\omega)$ in the 5-10 meV spectral range, and light-induced "superfluid density" extracted from a two-fluid model fit and expressed as a fraction of the total charge carrier density. Filled and open symbols indicate the results of the linear and square root reconstruction models respectively. The red dotted lines indicate the value of the corresponding quantity in equilibrium. This data was acquired at a base temperature T = 100 K with a fluence of approx. 18 mJ cm$^{-2}$ and a pump-pulse duration of $\Delta\tau_{pump} \approx 1$ ps. Transient optical spectra corresponding to these measurements are reported in supplementary figure S6.3.

Three spectrally-integrated figures of merit are extracted from the snapshots of $R(\omega,\tau)$, $\sigma_1(\omega,\tau)$ and $\sigma_2(\omega,\tau)$, and plotted as a function of pump-probe time delay $\tau$ in Fig. 2e, showing the time evolution of the system. The first two quantities are the frequency-averaged values of the reflectivity and $\sigma_1(\omega)$ from 5-10 meV, a frequency range which lies below the photo-induced energy gap, for which a zero-temperature superconductor with infinite lifetime would give values of 1 and 0 respectively. The third figure of merit is the fractional superfluid density which is proportional to the divergence of $\sigma_2(\omega)$. This is determined by fitting the photoexcited optical properties with a two-fluid model where one fluid represents the remaining normal carriers with a finite scattering rate and the other has zero scattering rate, giving a superconducting-like contribution. Details of this fitting procedure are given in supplementary section S7.

For low excitation fluences the system becomes superconducting-like after photoexcitation, and relaxes on a time scale of a few picoseconds. As already seen in the spectrally resolved measurements, for high excitation fluences the system enters a metastable regime in which the superconducting-like optical properties persist for much longer times, up to several nanoseconds.

We note that the temperature dependence reported in Ref. 12 shows transient superconducting-like optical properties up to a temperature of 150-200 K. For higher temperatures the gapping and extracted superfluid density are severely reduced. Examples of such spectra measured at room temperature are shown in Fig. 2d. Nevertheless, the pressure scaling reported in Ref. 13 suggests that traces of non-equilibrium superconductivity may survive up to higher temperatures, raising the prospect that with more effective driving a full manifestation of the metastable superconducting-like state may be possible at 300 K. To date, these experiments have been limited to excitation photon energies between 80 and 165 meV (20-40 THz), such that a more comprehensive search for a dominant excitation frequency scale has remained out of reach. Many potentially important resonances at lower frequencies ($h\nu < 80$ meV) have remained unexplored, primarily due to the lack of a suitable high-intensity pump source that operates in this range. In the present work, we explore excitation at energies between 24 and 80 meV (6-20 THz). This energy range hosts a number of excitations, both vibrational (phonons) and electronic in nature, including a broad polaronic peak seen in $\sigma_1$ centered at approximately 60 meV (15 THz). The possible relevance of this excitation has been highlighted in Ref. 22, although this prediction could not be tested to date.

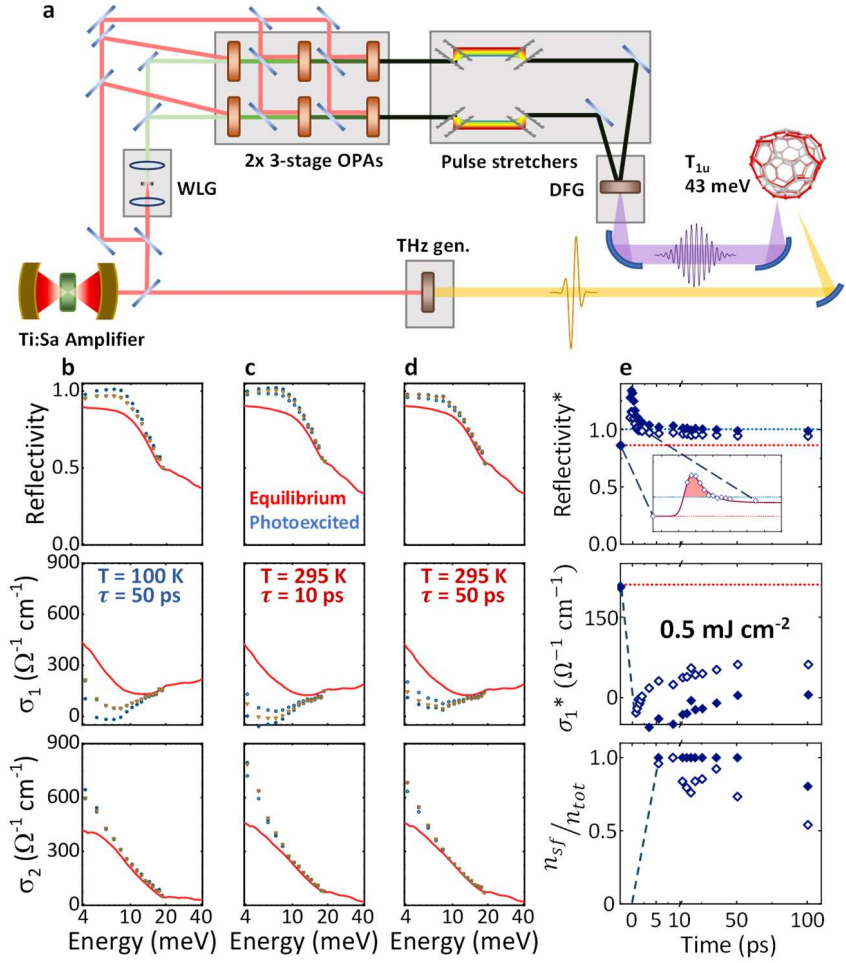

**Figure 3: Photo-induced metastable superconductivity in $K_3C_{60}$ generated with 41 meV excitation pulses. (a)** Schematic of the experimental set-up. Pump pulses with 41 meV (10 THz) photon energy (close to resonance with the illustrated phonon mode) are generated using a twin optical parametric amplifier (OPA) through subsequent chirped-pulse difference frequency generation (DFG) of the two stretched signal beams. The photoinduced changes in the far-infrared optical properties of $K_3C_{60}$ are detected with phase-sensitive transient THz time-domain spectroscopy. **(b)** Reflectivity (sample–diamond interface), real and imaginary part of the optical conductivity of $K_3C_{60}$ measured in equilibrium (red lines) and 50 ps after excitation (symbols). The data in filled blue circles, open blue circles and open brown triangles was obtained accounting for the inhomogeneous excitation of the probed volume under the assumption of a linear, square root and saturating (respectively) fluence dependence of the photo-induced changes in the complex refractive index of the material (Supplementary Section S6). This data was acquired at base temperature T = 100 K with pump pulses tuned to 41 meV (10 THz) center frequency and excitation fluence of approx. 0.4 mJ cm$^{-2}$. **(c)** Same quantities as in (b) but measured 10 ps after photoexcitation at base temperature T = 295 K. **(d)** Same quantities as in (c) but measured 50 ps after photoexcitation. **(e)** Time dependence of the average value of reflectivity and real part of the optical conductivity $\sigma_1(\omega)$ in the 5-10meV spectral range, and light-induced "superfluid density" extracted from a two-fluid model fit and expressed as a fraction of the total charge carrier density. Filled and open symbols indicate the results of the linear and square root reconstruction models respectively. The inset in the top panel highlights the early time delays for which light amplification ($R > 1$) is observed (red shading). The red dotted lines indicate the value of the corresponding quantity in equilibrium. This data was acquired at a base temperature T = 100 K with pump pulses tuned to 45 meV (11 THz) photon energy and excitation fluence of approx. 0.5 mJ/cm$^2$. Frequency-resolved spectra corresponding to these measurements can be seen in supplementary figure S6.4.

To achieve wide tuneability, we made use of a terahertz source based on chirped pulse difference frequency generation, mixing the near-infrared signal beams of two phase-locked optical parametric amplifiers[23]. This source, illustrated schematically in Fig. 3a and described in detail in supplementary section S4, was used to generate narrow-bandwidth pulses with photon energies spanning the range from 24 to 145 meV (6-35 THz). All measurements reported here were carried out with an excitation bandwidth of $h\Delta\nu_{\text{pump}} \approx$ 4 meV (1 THz) and $\Delta\tau_{\text{pump}} \approx 600$ fs pulse duration. The same probing protocol as that reported in Fig. 2 was utilized here to detect changes in the complex optical properties for probe energies spanning 4-18 meV (1-4.5 THz).

Figures 3b-d show reflectivity and complex conductivity spectra measured after photoexcitation with pulses tuned to 41 meV photon energy ($\lambda \approx 30$ μm, $f \approx 10$ THz) at base temperatures of 100 K and room temperature, respectively. The response is very similar to the case reported in Fig. 2 for 170 meV (41 THz) excitation, manifested on metastable timescales but persisting here up to room temperature – despite an almost two orders of magnitude weaker excitation fluence. Figure 3e displays the time-evolution of the optical properties, with a transient amplifying state at early pump-probe delays, as already reported in Ref. 24, that relaxes into a metastable state with superconducting-like optical properties.

Figure 4a shows the scaling with fluence of $R(\omega)$ and $\sigma_1(\omega)$ (averaged in the 5-10 meV range), as well as the fractional superfluid density in response to photoexcitation at 170 meV (41 THz) and 41 meV (10 THz). These measurements were carried out at a pump-probe time delay of 10 ps, and thus refer to the metastable phase. The figure shows how all figures of merit approach their equilibrium superconducting-state values as the fluence increases, with the fluence required being approximately 50 times less for 41 meV (10 THz) compared to 170 meV (41 THz) excitation.

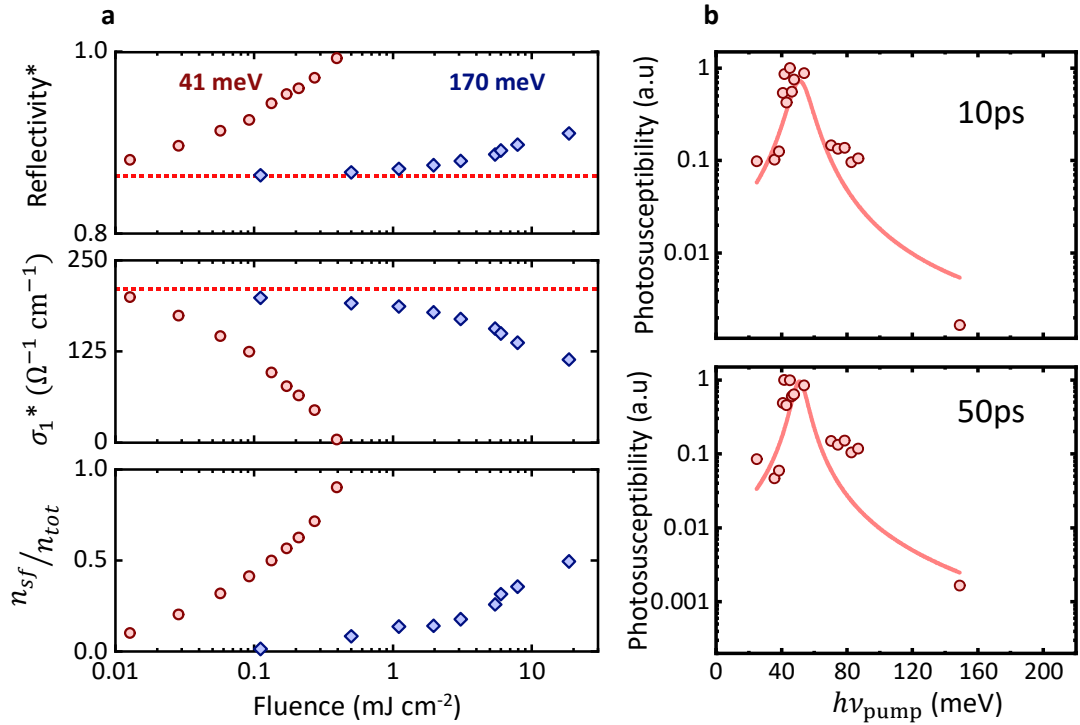

**Figure 4: Scaling of the out-of-equilibrium features of photo-induced metastable superconductivity in $K_3C_{60}$. (a)** Fluence dependence of the average value of the reflectivity and real part of the optical conductivity $\sigma_1(\omega)$ in the 5-10meV spectral range, and light-induced "superfluid density" extracted from a two-fluid model fit and expressed as a fraction of the total charge carrier density. Red and blue symbols indicate measurements with excitation pulses tuned to 41 meV (10 THz) and 170 meV (41 THz) central frequency. The red dotted lines indicate the value of the corresponding quantity at equilibrium. These data were acquired at a base temperature T = 100 K, at a time-delay Δt = 10ps, and with a pump pulse duration of $\Delta\tau_{pump} \approx$ 600fs. **(b)** Frequency dependence of the photo-susceptibility of $K_3C_{60}$ defined as the gradient of the lost spectral weight in $\sigma_1$ in the low-fluence limit (Supplementary Section S8) measured 10 ps and 50 ps after photo-excitation. These measurements were carried out at a base temperature T = 100 K. This data was obtained by accounting for the inhomogeneous excitation of the probed volume under the assumption of a square root fluence dependence of the photo-induced changes in the complex refractive index of the material. The solid curves are guides to the eye.

Similar fluence dependence measurements were carried out by varying the photon energy of the pump and maintaining a constant 4 meV (1 THz) bandwidth with 600 fs pulse duration. For all excitation photon energies between 24 meV (6 THz) and 145 meV (35 THz) the photoinduced changes in the optical properties were qualitatively similar to those shown in Figs. 2, 3 with only the size of the response for a given fluence differing. From each fluence dependence we extracted a figure of merit for the photo-susceptibility, defined as the rate of depletion of spectral weight in $\sigma_1$ with excitation fluence in the limit of low fluence (see supplementary section S8). Plots of the pump-frequency-dependent photo-susceptibility are shown in Fig. 4b for both 10 ps and 50 ps pump-probe time delay. A peak centered at 41 meV (10 THz) with approximately 16 meV FWHM width is observed in these measurements. In the next section we will discuss three distinct energy scales which coincide with this resonance, sequentially these relate to "on-ball" orbital excitations, phonons and finally excitons.

Superconductivity in alkali-doped fullerides is believed to be mediated by a dynamical Jahn-Teller distortion, which leads to an effective negative Hund's coupling for the orbitals of a single buckyball[25] and to a low spin S=1/2 state. A theoretical model based on this assumption has been successful at providing a quantitatively correct phase diagram for fulleride superconductors, based on ab-initio calculations[26,27]. Within this model, the local ground state of the system is a six-fold degenerate low-spin state, which features intra-orbital pairs that de-localize over two molecular orbitals. As detailed in supplementary section S10, a first set of local excited states also features such pairs, albeit with a different angular momentum (i.e. a different inter-orbital phase for the delocalized pair). Ab-initio calculations predict an energy splitting of 37 meV between these two sets of states[26,27]. The observed resonance may therefore be related to the creation of interorbital pairs with local angular momentum, which may also contribute to superconductivity, as suggested in the Suhl-Kondo mechanism[28-30]. However, it is not yet clear how exactly this excitation is transformed in the presence of tunneling between neighboring $C_{60}$ molecules, and why the creation of such pairs may support metastable superconductivity at such high temperatures. Furthermore, as the local parity of this excited state would be different from that of the ground state, condensation in this configuration may give rise to a superconductor with different symmetry. This possibility, whilst tantalizing, remains speculative and should be tested with more comprehensive ultrafast probing methods.

Turning to phonon excitations, we also note that the 41 meV resonance frequency identified here coincides with an infrared-active $T_{1u}$ phonon which predominantly consists of intramolecular motion of the C atoms. While the atomic motions of the 170 meV molecular mode discussed previously in Ref. 12 are directed along the tangential directions of the $C_{60}^{3-}$ molecule, those of the 41 meV mode are predominantly along the radial directions (see supplementary section S9). By performing frozen phonon calculations using density functional theory (DFT) we evaluated the different impact of these distortions on the three $t_{1u}$ molecular levels at the Fermi energy, which we map out from DFT wave functions as maximally-localized Wannier functions (supplementary section S9). In the undistorted $C_{60}^{3-}$ structure these molecular levels are degenerate. Applying a distortion along a $T_{1u}$ coordinate lifts this degeneracy leaving a doubly degenerate $t_{1u}$ orbital lowered in energy. This electronic configuration is prone to developing a Jahn-Teller distortion that may lead to an enhanced negative Hund's coupling, possibly facilitating the onset of superconductivity at higher temperatures. The strength of the induced splitting is quadratic in the phonon coordinate and is more significant for when driving the 41 meV mode compared to the 170 meV one, suggesting that the observed resonance may arise from a more efficient manipulation of the electronic degrees of freedom when driving the 41 meV $T_{1u}$ mode.

Finally, we address the electronic excitations discussed in Ref. 22, where an enhanced effect upon tuning the drive to lower frequencies was already predicted. This work proposed the existence of a dressed exciton in the excitation spectrum of $K_3C_{60}$ at the same energy scale as the resonance reported here. Excitation of the system at this frequency would generate excitons, that would act as a reservoir able to incoherently cool the quasiparticles, resulting in the re-emergence of superconductivity at base temperatures in excess of the equilibrium $T_c$.

We expect the significance of this discovery to be capitalized upon in future work. The extreme efficiency improvement due to resonant enhancement, nearing two orders of magnitude, is expected to also dramatically reduce unwanted dissipation. This, taken in conjunction with the observed nanosecond-long lifetime suggests that excitation of the sample with a train of pulses of only 400 µJ/cm² delivered at 100 MHz repetition rate – as determined by the inverse lifetime of this state - may yield continuous wave operation. Because this effect is documented here to persist up to room temperature, continuous wave operation would likely have important practical implications. To make this regime experimentally accessible, single order-of-magnitude improvements in the efficiency of the process, or in the light matter coupling strength, combined with suitable developments in high repetition rate THz sources would be required.

## Acknowledgments

The research leading to these results received funding from the European Research Council under the European Union's Seventh Framework Programme (FP7/2007-2013)/ERC Grant Agreement No. 319286 (QMAC, A.C.). We acknowledge support from the Deutsche Forschungsgemeinschaft (DFG) via the Cluster of Excellence 'The Hamburg Centre for Ultrafast Imaging' (EXC 1074 – project ID 194651731, A.C.). We thank Michael Volkmann and Peter Licht for their technical assistance. We are also grateful to Boris Fiedler and Birger Höhling for their support in the fabrication of the electronic devices used on the measurement setup, and to Jörg Harms for assistance with graphics.

# Supplementary Information

# S1. Sample growth and characterization

The K$_3$C$_{60}$ powder pellets used in this work were prepared and characterized as reported previously[1-3]. Stoichiometric amounts of ground C$_{60}$ powder and potassium were placed in a sealed pyrex vial, which was evacuated to a pressure of 10$^{-6}$ mbar. Whilst keeping the C$_{60}$ powder and solid potassium separated, the vial was kept at 523 K for 72 h and then at 623 K for 28 h such that the C$_{60}$ powder was exposed to pure potassium vapor. The vial was then opened inside an Ar glovebox (<0.1 ppm O$_2$ and H$_2$O), where the powder was reground and pelletized before annealing at 623K for 5 days. X-ray diffraction measurements were then carried out on the resulting K$_3$C$_{60}$ powder, which confirmed that it was phase pure, with an average grain size ranging between 100 and 400 nm. The static superconducting transition temperature was measured to be 19.8 K (in agreement with literature values) via magnetic susceptibility measurements upon zero field cooling and cooling in field with a field strength of 400 A/m.

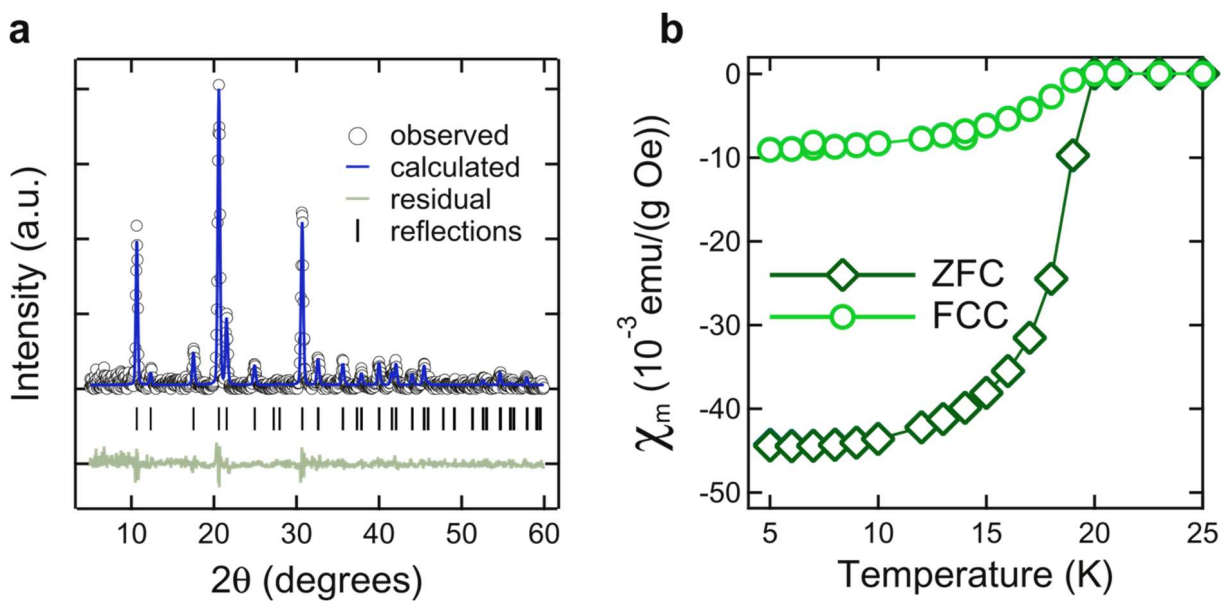

**Figure S1.1: a.** X-ray diffraction data and single f.c.c. phase Rietveld refinement for the K$_3$C$_{60}$ powder used in this work. **b.** Temperature dependence of the sample magnetic susceptibility measured by SQUID magnetometry upon cooling without (ZFC: zero field cooling) and with a magnetic field applied (FCC: field cooled cooling).

## S2. Determination of the equilibrium optical properties

The equilibrium reflectivity was measured for photon energies between 5 meV and 500 meV using a commercial Fourier-transform infrared spectrometer (FTIR) equipped with a microscope at the SISSI beamline in the Elettra Synchrotron Facility (Trieste, Italy), as reported previously[1-3]. The sample was pressed by a diamond window into a sealed holder in order to obtain an optically flat interface and prevent exposure to air. This procedure was carried out inside an Ar filled glove box (<0.1 ppm $O_2$ and $H_2O$) before the sealed sample was removed and mounted on a He cooled cryostat to enable temperature dependent measurements. The $K_3C_{60}$ reflectivity spectra were referenced against a gold mirror placed at the sample position.

In order to extract the complex optical conductivity a Kramers-Kronig algorithm for samples in contact with a transparent window[4] was used. This requires data at all frequencies, which were obtained, at low energies (<5 meV) using an extrapolation based on a Drude-Lorentz fit, and at high energies (>500 meV) using data measured on single crystal samples reported in Refs. 5,6.

The equilibrium properties are shown in figure S2.1 for temperatures of 100 K and 300 K. This and further data measured at different temperatures and pressures were already reported in Refs. 1,2 and discussed also in comparison with data obtained from single crystals.

These data were fitted with a Drude-Lorentz model, which is given by the following equation:

$$\sigma_1(\omega) + i\sigma_2(\omega) = \frac{\omega_p^2}{4\pi}\frac{1}{\gamma_D - i\omega} + \frac{\omega_{p,osc}^2}{4\pi}\frac{\omega}{i(\omega_{0,osc}^2 - \omega^2) + \gamma_{osc}\omega}$$

Here the first term represents the Drude response of the free carriers with $\omega_p$ and $\gamma_D$ representing the plasma frequency and scattering rate respectively, whereas the second term captures the mid infrared absorption in the form of a Lorentz oscillator centered at frequency $\omega_{0,osc}$ with plasma frequency $\omega_{p,osc}$ and damping rate $\gamma_{osc}$. The equilibrium data reported here was used to normalize the transient optical spectra of $K_3C_{60}$ measured upon photoexcitation, as discussed in detail in section S6.

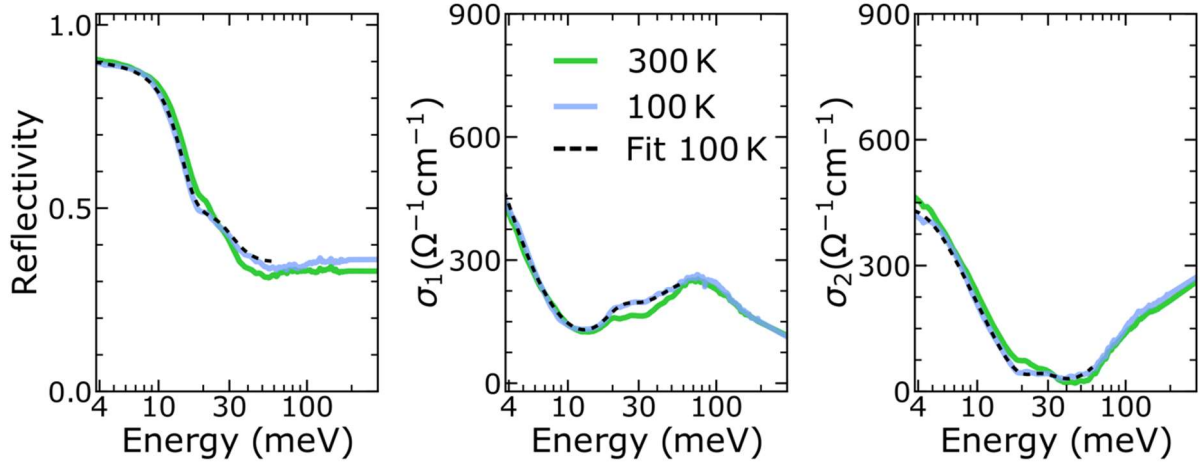

**Figure S2.1:** Equilibrium optical properties (reflectivity, real, and imaginary part of the optical conductivity) of $K_3C_{60}$ measured at a temperature of 100 K (blue) and 300 K (green). The black dashed curve is a Drude-Lorentz fit to the optical conductivity at 100 K in the range from 3 meV to 60 meV as described in the text.

## S3. High fluence mid-infrared source

For the data reported in figure 2 and in figure 4(a) at 170 meV (41 THz) excitation, the pump pulses were generated via difference frequency mixing (DFG) of the signal and idler output of a three-stage home-built optical parametric amplifier (OPA). A commercial Ti:$Al_2O_3$ amplifier delivering 60 fs duration pulses at 800 nm central wavelength was used to drive the OPA, and the DFG process was performed using a 0.5 mm thick GaSe crystal, resulting in ~100 fs long pulses. The 170 meV pulses were then propagated through a highly dispersive 16 mm long $CaF_2$ rod, stretching their duration to ~1 ps. The spectrum of the pump pulses was characterized using a home built FTIR spectrometer. Their duration was measured by cross-correlation with a synchronized, 35 fs long, 800 nm wavelength pulse in a 50 μm thick GaSe crystal. While a certain degree of tunability is also given by this source, its useful operation range spans between 80 and 320 meV, hence it was only used for the high-intensity experiments at 170 meV excitation.

## S4. Frequency-tunable narrowband terahertz and mid-infrared source

For the experiments that required tunability of the excitation pulses down to the THz gap, a different source was used. This source is based on the principle of chirped-pulse difference frequency generation (CP-DFG) in organic non-linear optical crystals, namely DAST and DSTMS of approximately 600 μm thickness. The principle of operation of this new source is described in detail in Ref. 7. A commercial Ti:Al$_2$O$_3$ amplifier is used to drive two identical three-stage OPAs which are seeded by the same white-light, such that the signal beams have the same phase-fluctuations. The ~100 fs signal pulses are then chirped using a pair of transmission-grating-based stretchers as depicted in figure 3(a). This arrangement enables continuous tuning of the pulse durations by varying the distance between the gratings in each pair, effectively enabling continuous tuning of the pump-pulse bandwidth. For this experiment the pump pulse bandwidth was kept constant at 4 meV by maintaining a signal pulse duration of ~600 fs, as measured using a home-built second harmonic-based Frequency-Resolved-Optical-Gating (FROG) device. Frequency tuning of the generated excitation pulses was carried out both by varying the central wavelengths of the two OPA signal beams, and by varying the time delay between the chirped signal pulses in the DFG crystal (for fine tuning). For each measurement the pump frequency spectrum was measured via FTIR (Fourier Transform Infrared Spectroscopy).

## S5. Measurements of the transient THz reflectivity

The experiments presented in Figures 2, 3, and 4 were performed on compacted K$_3$C$_{60}$ powder pellets pressed against a diamond window to ensure an optically flat interface. As K$_3$C$_{60}$ is water and oxygen sensitive, the pellets were sealed in an air-tight holder and all sample handling operations were performed in an Argon filled glove box with <0.1 ppm O$_2$ and H$_2$O. The sample holder was then installed at the end of a commercial Helium cold-finger (base temperature 5K), to cool the pellets down to a temperature of 100 K.

The changes in the properties of the sample following photoexcitation were measured using time-domain THz-spectroscopy.

The mid-infrared pump induced changes in the low frequency optical properties, were retrieved using transient THz time domain spectroscopy in two different experimental setups. The THz probe pulses were generated via optical rectification in a 0.2 mm thick (110)-cut GaP crystal starting from 800 nm pulses with a duration of ~80 fs and 35 fs, respectively. Whilst in one setup these 800 nm were derived from the same laser used for pumping the source described in section S4, the 35 fs, 800 nm pulses were generated by a second Ti:Al$_2$O$_3$ amplifier optically synchronized to that used to pump the high-intensity mid-infrared source described in section S3. The THz probe pulses were then focused onto the sample with incidence angles of 30 and 0 degrees, respectively. After reflection from the sample, the electric field profile of the THz pulses was reconstructed in a standard electro-optic sampling setup, using a (110)-cut 0.2 mm GaP crystal supported on a 1 mm thick (100)-cut GaP substrate to delay internal reflections. The setup combined with the frequency tunable narrowband source had a measurement bandwidth that extended between 4 and 18 meV, while the other spanned between 4 meV to 29 meV. The time resolution of both setups is determined by the measurement bandwidth and is ~250 fs and ~150 fs respectively. We note that while both setups were capable of generating and detecting frequencies below 4 meV, the lower frequency limit is set by our capabilities of focusing the probe THz pulses to diameters smaller than those of the pump beam. In the two experimental setups used in this work the pump beam was focused down to ~800 µm and ~450 µm respectively. These values are below the probe beam diameters measured in the two setups at 4 meV, ensuring that probe frequencies larger than 4 meV probed a homogeneously excited area.

To minimize the effects on the pump-probe time resolution due to the finite duration of the THz probe pulse, the experiments were performed as described in Refs. 8, 9. The pump-probe time delay was controlled by fixing the delay between the 800 nm gating pulse and the mid-infrared pump pulse $\tau$. The transient THz field was then obtained by scanning the delay $t$ relative to both.

In order to simultaneously retrieve both the 'pump on' ($E_{THz}^{on}(t,\tau)$) and 'pump off' ($E_{THz}^{off}(t)$) probe fields, a differential chopping scheme was deployed. The scheme was different for the two above mentioned setup. For the narrowband, frequency tunable setup which operated at a repetition rate of 1 kHz, the THz probe pulse was chopped at a frequency of 500 Hz and the mid-infrared pump pulse was chopped at ~ 357 Hz. The electro-optic sampling signal was then fed to two lock-in amplifiers reading out $V_{LIA1}$ at 500 Hz and $V_{LIA2}$ at 143 Hz respectively. For the high-intensity setup, operating at 2 kHz repetition rate, the THz probe pulse was chopped at a frequency of 1 kHz and the mid-infrared pump was chopped at 500 Hz. In this case, the electro-optic sampling signal was filtered by two lock-in amplifiers operating at 1 kHz and 500 Hz respectively. $E_{THz}^{off}(t)$ and $\Delta E_{THz}(t,\tau)$ were then extracted from the signals in the two lock-ins using the following formulas:

$$E_{THz}^{off}(t) = V_{LIA1}(t,\tau) - \alpha V_{LIA2}(t,\tau)$$

$$\Delta E_{THz}(t,\tau) = E_{THz}^{on}(t,\tau) - E_{THz}^{off}(t) = \alpha V_{LIA2}(t,\tau)$$

where $\alpha$ is a calibration constant determined experimentally on an InSb reference sample. This is done by extracting $\Delta E_{THz}(t,\tau)$ as the difference of two separate measurements of $E_{THz}^{on}(t,0)$ and $E_{THz}^{off}(t)$ performed with the first lock-in amplifier and by chopping only the THz probe pulse while leaving the mid-infrared pump pulse either always on or always off. Equating the value of $\Delta E_{THz}(t,\tau)$ determined in this way to the one with differential chopping yields the calibration constant.

## S6. Determination of the transient optical properties

From the measured changes in the reflected probe field (see section S5), the transient complex reflection coefficient of the sample $\tilde{r}(\omega,\tau)$ can be determined by taking the Fourier transform along t of both $E_{THz}^{off}(t)$ and $\Delta E_{THz}(t,\tau)$ and using the following equation:

$$\frac{\Delta \tilde{E}_{THz}(\omega,\tau)}{\tilde{E}_{THz}^{off}(\omega)} = \frac{\tilde{r}(\omega,\tau) - \tilde{r}_0(\omega)}{\tilde{r}_0(\omega)}$$

where $\tilde{r}_0(\omega)$ is the equilibrium complex reflection coefficient, obtained as described in section S2. In the cases where the pump light penetrates significantly further into the sample than the probe light, one can assume that the probe pulse samples a volume in the material that has been homogeneously excited by the pump. In this case, it is possible to directly extract the complex-valued optical response functions by inverting the Fresnel equations. By applying this procedure to the data measured at 41 meV (10 THz) pump photon energy, we obtain the spectra shown in figure S6.1.

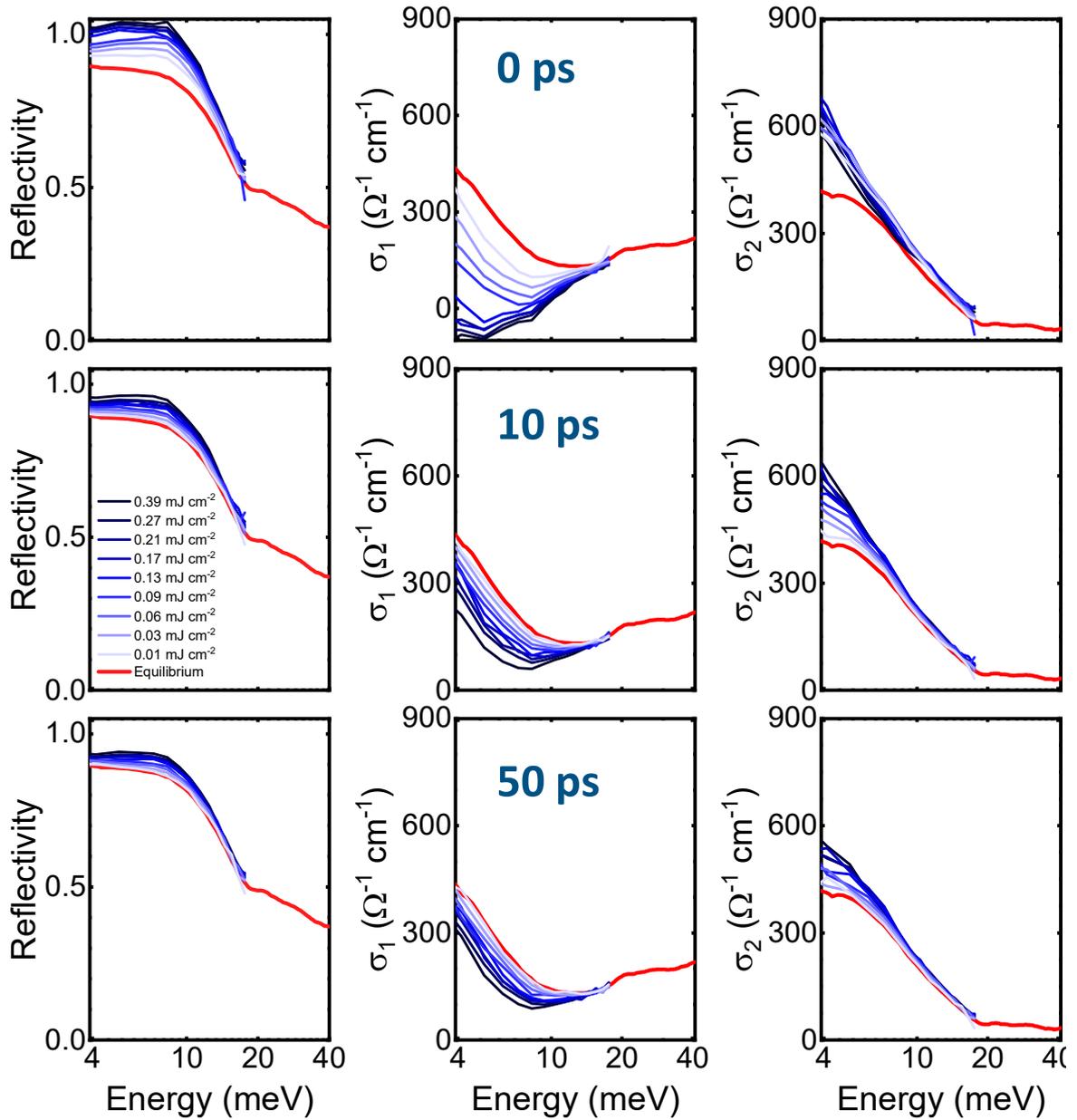

**Figure S6.1: Raw transient optical spectra at 41 meV (10 THz) pump photon energy.** Reflectivity (sample-diamond interface), real, and imaginary parts of the optical conductivity measured at equilibrium (red lines) and after photoexcitation (blue lines) at increasing pump-probe time delays and excitation fluences indicated in the figure. The data in this figure underestimates the changes in the optical properties, as the mismatch in penetration depth between the pump and probe fields is neglected.

This procedure is not correct for the case of $K_3C_{60}$, where the penetration depth of the probe (~600-900 nm) exceeds that of the pump (~500 nm at 10 THz, ~200 nm at 41 THz), such that the probe interrogates an inhomogeneously excited volume (Figure S6.2(a)).

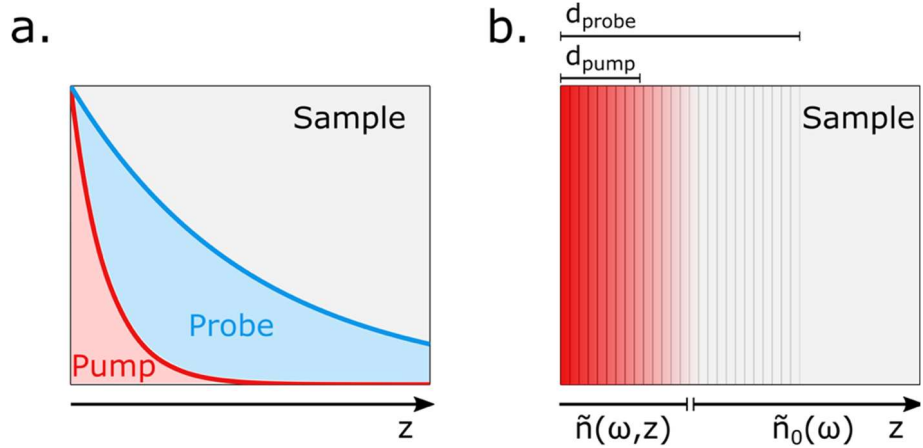

**Figure S6.2: a.** Schematics of pump-probe penetration depth mismatch. **b.** Multi-layer model with exponential decay used to calculate the pump-induced changes in the complex refractive index $\tilde{n}(\omega,\tau)$ for each pump-probe delay $\tau$. The transition from red to background (grey) represents the decaying pump-induced changes in $\tilde{n}(\omega,z)$.

As the pump penetrates into the material, its intensity is reduced, and it will induce progressively weaker changes in the refractive index of the sample. This situation is modeled by "slicing" the probed thickness of the material into thin layers (figure S6.2(b)), where we assume that the pump-induced changes in the refractive index $\Delta\tilde{n}$ scale according to the pump intensity in the layer, i.e. $\tilde{n}(\omega,z,\tau) = \tilde{n}_0(\omega) + \Delta\tilde{n}(\omega,\tau,I(z))$. The pump intensity $I(z)$ is assumed to follow the dependence $I(z) = I_0 e^{-z/d_{pump}}$, where $d_{pump} = \lambda_{pump}/4\pi Im\left(n_0(\omega_{pump})\right)$. Here, the refractive index of the material at the pump frequency, $n_0(\omega_{pump})$ is taken to be the one at equilibrium. Additionally, an assumption is made on the functional form for the dependence of $\Delta\tilde{n}$ on the pump intensity. Here, we consider two different forms given by:

(1) $\quad \Delta\tilde{n}(\omega,\tau,z) \propto I(z)$

(2) $\quad \Delta\tilde{n}(\omega,\tau,z) \propto \sqrt{I(z)}$

Respectively, these equations result in the following depth-dependent functional forms for the spatial profile of the refractive index:

(1) $\quad \tilde{n}(z,\omega,\tau) = \tilde{n}_0(\omega) + \Delta\tilde{n}(\omega,\tau)e^{-z/d_{pump}}$

(2) $\quad \tilde{n}(z,\omega,\tau) = \tilde{n}_0(\omega) + \Delta\tilde{n}(\omega,\tau)e^{-z/2d_{pump}}$

where $\Delta\tilde{n}(\omega,\tau)$ represents the pump-induced change in the refractive index of the material at the sample surface.

For each time delay $\tau$ and probe frequency $\omega_i$, the complex reflection coefficient $\tilde{r}(\Delta\tilde{n})$ of the multilayer stack described above is calculated using the transfer matrix method[10], keeping $\Delta \tilde{n}$ as a free parameter. To numerically extract the value of $\Delta \tilde{n}(\omega, \tau)$ we minimize the following function:

$$\left| \frac{\Delta \tilde{E}_{THz}(\omega_i)}{\tilde{E}_{THz}^{off}(\omega_i)} - \frac{\tilde{r}(\omega_i, \Delta n) - \tilde{r}_0(\omega_i)}{\tilde{r}_0(\omega_i)} \right|$$

By then taking $\tilde{n}(\omega, \tau) = \tilde{n}_0(\omega) + \Delta \tilde{n}(\omega, \tau)$, one obtains the refractive index of the material as if it had been homogeneously excited. From $\tilde{n}(\omega, \tau)$ we then calculate $R(\omega, \tau)$, $\sigma_1(\omega, \tau)$ and $\sigma_2(\omega, \tau)$ as plotted in the main text. Figures S6.3 and S6.4 display extended data sets measured at increasing pump-probe delays with pump photon energies of 170 meV (41 THz) and 45 meV (11 THz) respectively. Therein we report reflectivity (sample-diamond interface), real and imaginary parts of the optical conductivity after reconstruction under the assumptions of models (1) and (2), identified with hollow and filled circles respectively.

At early delays, for both excitation mechanisms and reconstruction assumptions, the reconstructed reflectivity is higher than one, and the real part of the optical conductivity is negative, indicative of amplification of the incoming THz probe radiation, as discussed previously in Ref. 11. In all cases, this non-equilibrium driven state then relaxes into a superconducting-like state with a fully gapped $\sigma_1(\omega)$ and a divergence $\propto 1/\omega$ in the $\sigma_2(\omega)$ spectrum. At even later delays the optical spectra are those of a finite temperature superconductor. These optical properties can be interpreted in the context of a two fluid model, in which a varying density of uncondensed quasi-particles also contributes to the terahertz response as discussed in Ref. 12.

It has been pointed out in Ref. 13 that if the photo-induced changes in the sample saturate with increasing fluence, the profile of $\Delta \tilde{n}(\omega, z)$ will be deformed into a non-exponential shape. In that work, the authors suggest assuming the following functional form for the change in conductivity with fluence:

$$\Delta \tilde{\sigma}(\omega, \tau, z; I_{sat}) \propto \frac{I(z)/I_{sat}}{1 + I(z)/I_{sat}}$$

where $I(z) = I(z=0)\, e^{-z/d_{pump}}$ is the local excitation fluence and $I_{sat}$ is the saturation fluence which must be determined experimentally from a fluence dependence measurement.

In the following, we benchmark this additional reconstruction model against the ones described above by applying it to the dataset measured with a pump frequency of 10 THz and fluence 0.4 mJ cm$^{-2}$ at the three time delays $\tau = 0$ ps, 10 ps, and 50 ps for which a fluence dependent measurement was available.

Analogously to the procedure used previously with models (1) and (2), we construct the transfer matrix of a multi-layer stack where now the depth dependent complex conductivity has the profile given by (3). Note that in the treatment using model (3) we cast all the equations in terms of $\tilde{\sigma}$ instead of $\tilde{n}$ to be consistent with Ref. 13:

$$(3) \quad \tilde{\sigma}(z, \omega, \tau; I_{sat}) = \tilde{\sigma}_0(\omega) + \Delta\tilde{\sigma}(\omega, \tau) \frac{I(z)/I_{sat}}{1 + I(z)/I_{sat}} \frac{(1 + I(0)/I_{sat})}{I(0)/I_{sat}}$$

As with models (1) and (2) above, we can now extract the complex reflection coefficient $\tilde{r}(\omega, \Delta\tilde{\sigma}; I_{sat})$ of the full multilayer stack and search for the value of $\Delta\tilde{\sigma}$ which minimizes the function:

$$\left| \frac{\Delta\tilde{E}_{THz}(\omega_i)}{\tilde{E}_{THz}^{off}(\omega_i)} - \frac{\tilde{r}(\omega_i, \Delta\tilde{\sigma}; I_{sat}) - \tilde{r}_0(\omega_i)}{\tilde{r}_0(\omega_i)} \right|$$

The optical properties resulting from this procedure are shown in figure S6.5 alongside those obtained with model (2). For the data measured at 0 ps time delay, model (2) and (3) show slight differences in the reconstructed optical spectra but both still show amplification of the probe light (R>1). At the longer time delays of 10 ps and 50 ps, the optical spectra show superconducting-like optical properties and the two models give virtually indistinguishable results.

To summarize, the time-evolution of K$_3$C$_{60}$ following photo-excitation is independent of the used reconstruction, and only the specific values of pump-probe delay up to which amplification, fully gapped superconductor, and finite temperature superconductor appear are affected by this choice.

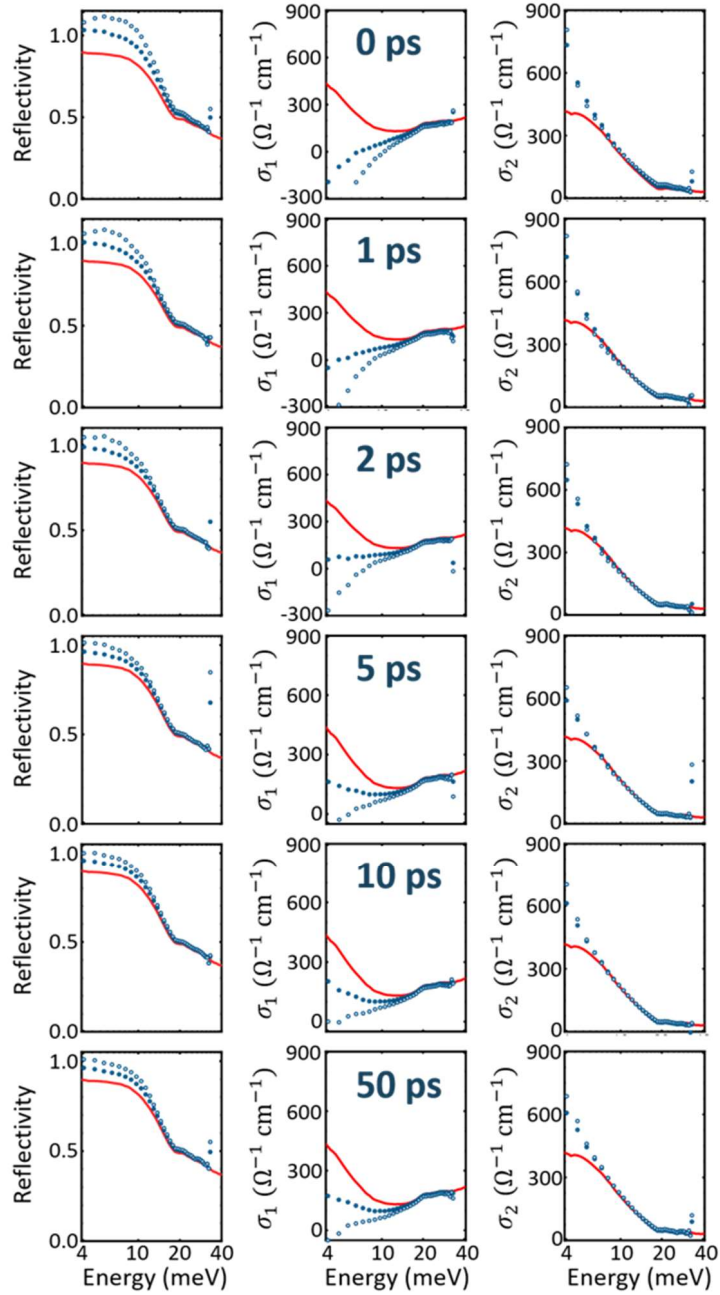

**Figure S6.3: Comparison of linear and square root reconstruction in the transient optical spectra at 170 meV (41 THz) pump-photon energy.** Reflectivity (sample-diamond interface), real, and imaginary parts of the optical conductivity measured at equilibrium (red lines) and after photoexcitation (blue symbols) at increasing pump-probe time delays indicated in the figure. The data in filled (open) symbols reconstructed under the assumption of a square-root (linear) fluence dependence of the changes in complex refractive index of the material. These data were measured at 18 mJ cm$^{-2}$ excitation fluence, and at a base temperature of 100 K.

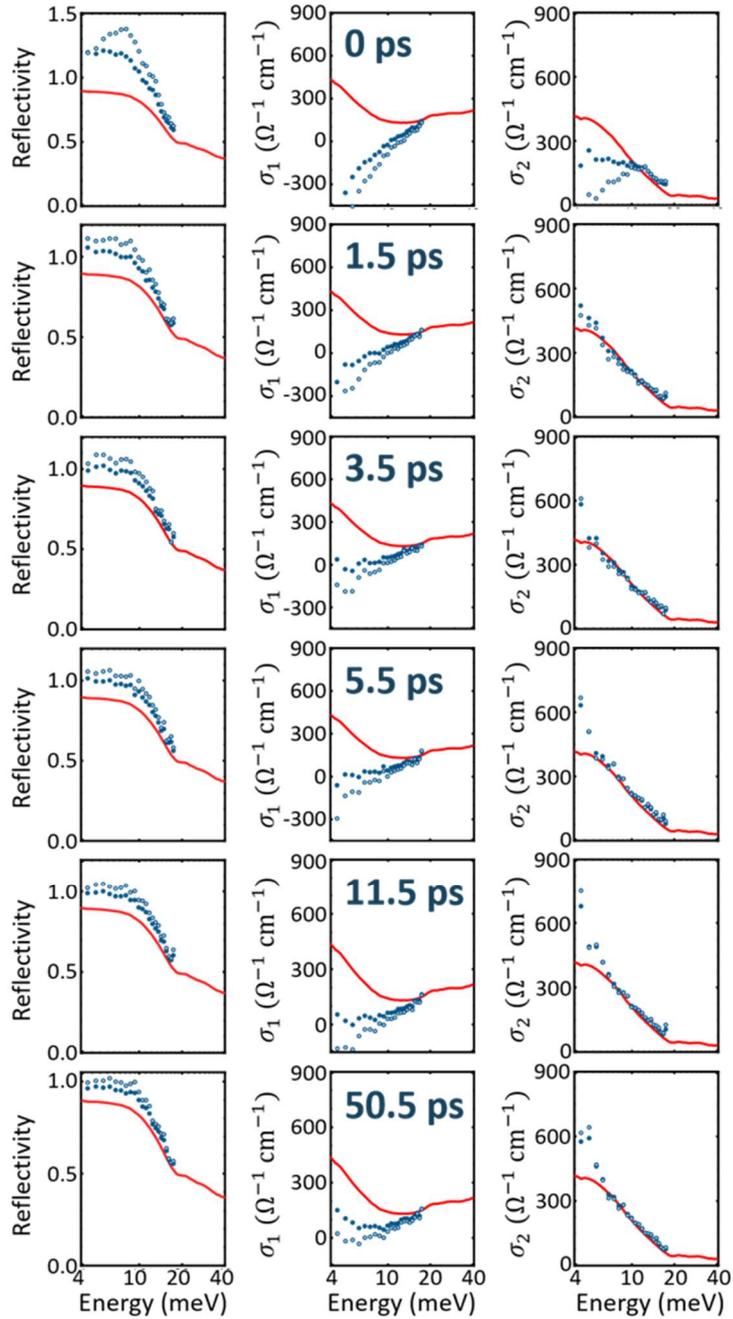

**Figure S6.4: Comparison of linear and square root reconstruction in the transient optical spectra at 45 meV (11 THz) pump-photon energy.** Reflectivity (sample-diamond interface), real, and imaginary parts of the optical conductivity measured at equilibrium (red lines) and after photoexcitation (blue symbols) at increasing pump-probe time delays indicated in the figure. The data in filled (open) symbols reconstructed under the assumption of a square-root (linear) fluence dependence of the changes in complex refractive index of the material. These data were measured at 0.5 mJ cm$^{-2}$ excitation fluence, and at a base temperature of 100 K.

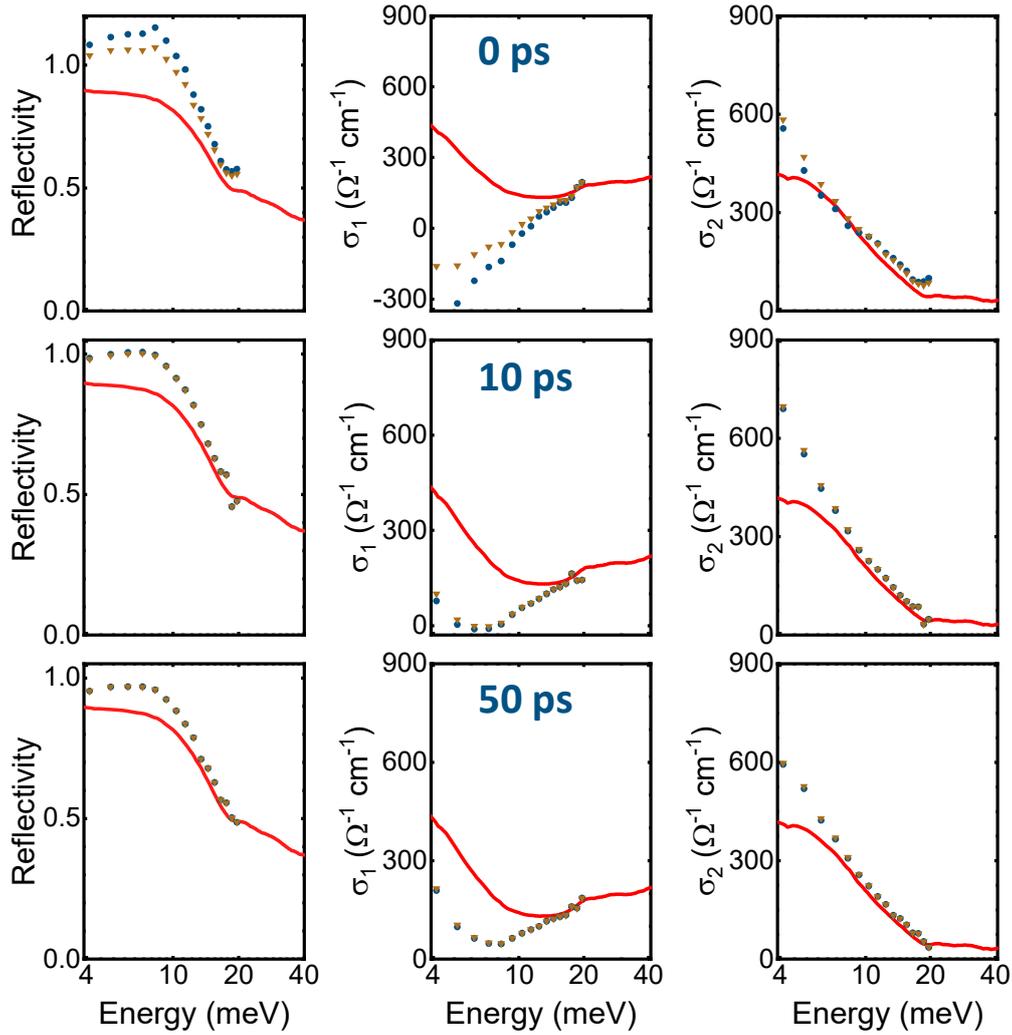

**Figure S6.5: Comparison of the transient optical spectra measured at 41 meV (10 THz) pump photon energy and reconstructed with a saturation and square root model.** Reflectivity (sample-diamond interface), real, and imaginary parts of the optical conductivity measured at equilibrium (red lines) and after photoexcitation at increasing pump-probe time delays indicated in the figure. The data in blue circles is reconstructed under the assumption of a square-root fluence dependence of the changes in complex refractive index, whereas the data in orange triangles is reconstructed under the assumption of a saturating fluence dependence of the changes in the complex conductivity. For the saturating model, the values $I_{sat} = 0.04, 0.20,$ and $0.22$ mJ cm$^{-2}$ were used for data taken at 0, 10 and 50 ps respectively. These data were measured at 0.4 mJ cm$^{-2}$ excitation fluence, at a base temperature of 100 K.

## S7. Fitting of the transient optical spectra

The transient optical conductivity spectra presented in figures 2-3 as well as for each fluence in figure 4 were fitted with a two-fluid model according to the following equation:

$$\tilde{\sigma}(\omega,\tau) = \frac{\pi}{2}\frac{\Lambda_s(\tau)\,e^2}{m}\delta[\omega=0] + i\frac{\Lambda_s(\tau)\,e^2}{m}\frac{1}{\omega}$$

$$+ \frac{\Lambda_n(\tau)\,e^2}{m}\frac{1}{\gamma_D - i\omega}$$

$$+ \sum_{n=1}^{2}\frac{B_n\omega}{i(\Omega_n^2 - \omega^2) + \gamma_n\omega}$$

Here the first term captures the frequency dependent contribution from the supercarriers with density $\Lambda_s$, the second term captures the Drude contribution of the normal carriers with density $\Lambda_n$ and scattering rate $\gamma_D$. Finally, we include a sum over two Lorentz

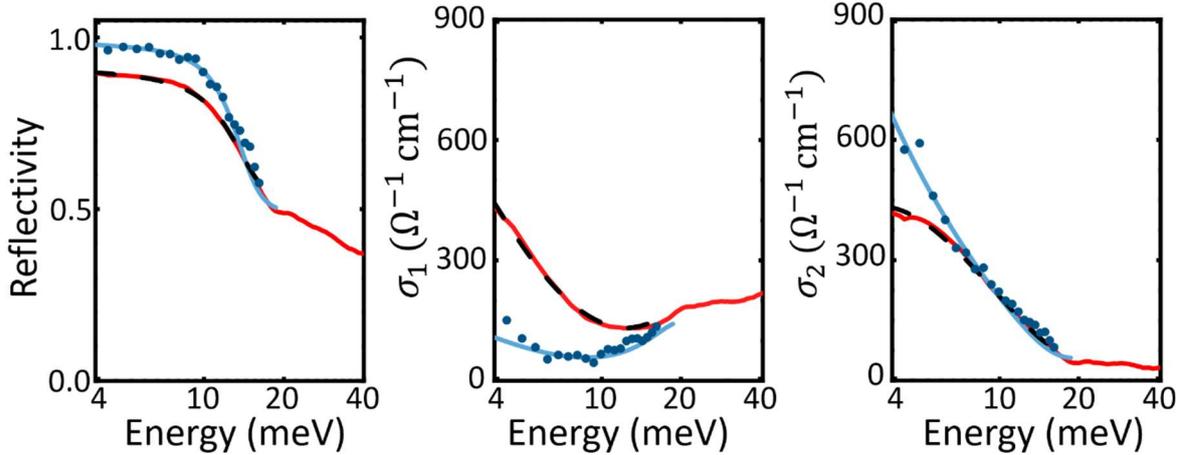

**Figure S7.1: Two-fluid fit to the transient spectrum.** Reflectivity, real ($\sigma_1$) and imaginary ($\sigma_2$) parts of the optical conductivity measured in equilibrium at 100 K (red) and 50 ps after photoexcitation with a fluence of 0.5 mJ cm$^{-2}$ at 45 meV (11 THz) photon energy. The fit to the equilibrium data using the procedure described in this section is shown as a dashed black line and gives zero superfluid density. The two-fluid fit to the transient data generated using the same procedure is shown as a solid blue line and returns a superfluid fraction $\Lambda_s/(\Lambda_n + \Lambda_s) = 73\%$. The data in this figure was reconstructed under the assumption of a square root dependence of the change in refractive index on excitation fluence (see supplementary section S6).

oscillators in order to capture the broad midinfrared absorption peak centered at around 60 meV.

The transient data are fitted at each delay $\tau$ using the parameter-set that captures the equilibrium optical conductivity spectra as a starting condition, and leaving only $\Lambda_s$ and $\Lambda_n$ free to vary, as though the effect of the pump is to simply convert carriers from the normal to the superconducting fluid.

Figure S7.1 shows representative fits to transient data measured at 100 K base temperature and at 50 ps time delay, as well as to the 100 K equilibrium spectra. Importantly, while the fit of the equilibrium data converges to a superfluid fraction $\Lambda_s/(\Lambda_n + \Lambda_s)$ which is equal to zero, the fit to the transient data yields $\Lambda_s/(\Lambda_n + \Lambda_s) = 0.73$. The transient optical data was fitted at each time delay and driving frequency, yielding the time and frequency dependence of the superfluid fractions shown in figures 2(e), 3(e), and 4(a).

## S8. Extracting the frequency-dependent photosusceptibility

In figure 4(b) we introduce a figure of merit, referred to as the 'photosusceptibility', which can be used to quantitatively compare the efficiency with which the metastable light-induced superconducting state is generated in K$_3$C$_{60}$ for different excitation frequencies. For each excitation photon energy, transient optical spectra were measured at different excitation fluences $\mathcal{F}$. From these fluence dependent spectra we extract the loss in spectral weight of $\sigma_1(\omega)$ after photoexcitation in the 5-10 meV spectral range, calculated as:

$$SWL(\mathcal{F}) = \int_{5\ \text{meV}/\hbar}^{10\ \text{meV}/\hbar} \sigma_1^{eq}(\omega) - \sigma_1^{photo}(\omega, \mathcal{F})\, d\omega$$

where $\sigma_1^{eq}(\omega)$ and $\sigma_1^{photo}(\omega, \mathcal{F})$ are the $\sigma_1(\omega)$ spectra measured in equilibrium and upon photoexcitation respectively. The $SWL(\mathcal{F})$ data is then fitted with the following phenomenological function:

$$A\left(\frac{1}{1 + Be^{-\frac{4B\mathcal{F}}{A}}} - \frac{1}{2}\right)$$

where $\mathcal{F}$ represents the excitation fluence and $A$, $B$ are free parameters. The 'photosusceptibility' plotted in figure 4(b) is equal to $B$, which is the gradient of this function evaluated at zero fluence. Figure S8.1 shows the fluence-dependent data and corresponding fit for one exemplary dataset.

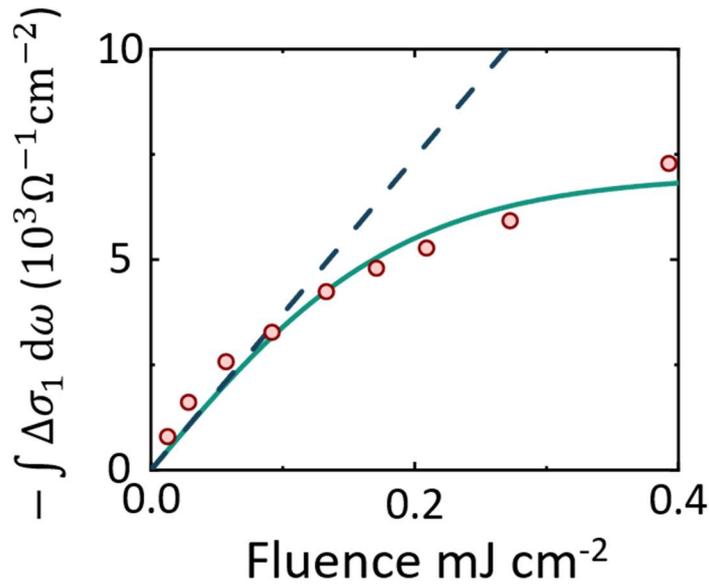

**Figure S8.1: Extracting photosusceptibility from the fluence-dependent data.** Lost spectral weight in the real part of the optical conductivity between 5 and 10 meV as a function of fluence (red circles), measured 10 ps after photoexcitation at 100 K with a pump spectrum centered at 41 meV (10 THz). The fit is shown as a solid green line, with the gradient at zero fluence (which we define as the photosusceptibility) shown as a dashed blue line. The data in this figure was reconstructed under the assumption of a square root dependence of the change in refractive index on excitation fluence (see supplementary section S6).

## S9. Density functional theory calculations

In this section, we address how the displacement of phonon modes affects the electronic properties of $K_3C_{60}$. Specifically, we consider the molecular orbitals and their response to the change in the crystal structure. To carry out this investigation, a first-principles approach based on density functional theory (DFT) was used. The starting point is the unit cell of $K_3C_{60}$ containing sixty carbon and three potassium atoms. Before computing the phonon spectrum, this unit cell is structurally relaxed, and the resulting lattice constants and atomic coordinates are listed in table S9.1.

Next, the phonon spectrum of $K_3C_{60}$ is computed from the force constant matrix utilizing a finite displacement approach[14]. In total, there are 186 non-translational phonon modes covering the symmetries of point group *m-3*. Specifically, there are 24 $T_u$, 7 $E_u$, 23 $T_g$, 8 $E_g$, and 8 $A_g$ modes. Note that only the modes of $T_u$ character are infrared active, and we list their computed frequencies in the table S9.2.

We utilized a frozen phonon approach to estimate the impact of these distortions on the molecular levels. Therefore, we modulated our equilibrium crystal structure with the eigen-displacements of these modes. We then created a low energy Hamiltonian for these structures by computing the maximally localized Wannier functions for the valence band electrons. Note that since the three valence bands are well separated in energy from other orbital-like bands our method does not require a disentanglement procedure.

| Lattice vectors | | | |
|---|---|---|---|
| a | 14.175 Å | Alpha | 90° |
| b | 14.175 Å | Beta | 90° |
| c | 14.175 Å | Gamma | 90° |
| Atomic positions according to Space Group 202 (Fm-3) | | | |
| Element | Wykoff label | X | y | c |
| C | H | 0.00000 | 0.54991 | 0.24682 |
| C | I | 0.58242 | 0.10057 | 0.21408 |
| C | I | 0.66275 | 0.05092 | 0.18294 |
| K | C | 0.25000 | 0.25000 | 0.25000 |
| K | A | 0.00000 | 0.00000 | 0.00000 |

**Table S9.1:** Structural parameters of $K_3C_{60}$ from first-principles computations

Our calculations focused on the three degenerate $t_{1u}$ molecular levels at the Fermi energy, which we mapped out from DFT wave functions as maximally-localized Wannier functions. In the equilibrium structure, the onsite energy of these molecular levels is degenerate; however, deforming the crystal by applying a $T_{1u}$ polar distortion lifts this degeneracy. Thereby, similar to a Jahn-Teller distortion, the symmetry breaking of the crystal structure splits the level into a double and a single degenerate orbital. For the 43.2 meV and 173.4 meV phonon modes, this splitting manifests as a lowering in the energy of the double degenerate orbital. A schematic visualization of this is depicted in the inset to figure S9.1(a). Diagrams illustrating the distortion of the $C_{60}$ molecule for the 43.2 meV and 173.4 meV modes (labelled 'A' and 'B' and corresponding to mode numbers 4 and 21 in table S9.2 respectively) are shown in figure S9.1(b).

| Number: | $h\nu_{pump}$ (meV) |
|---|---|
| 1 | 2.2 |
| 2 | 14.1 |
| 3 | 42.4 |
| 4 (A) | 43.2 |
| 5 | 48.3 |
| 6 | 60.5 |
| 7 | 62.6 |
| 8 | 71.5 |
| 9 | 80.6 |
| 10 | 83.9 |
| 11 | 85.9 |
| 12 | 91.1 |
| 13 | 92.0 |
| 14 | 118.6 |
| 15 | 122.8 |
| 16 | 147.5 |
| 17 | 148.3 |
| 18 | 149.8 |
| 19 | 163.7 |
| 20 | 165.4 |
| 21 (B) | 173.4 |
| 22 | 176.9 |
| 23 | 184.8 |
| 24 | 185.3 |

**Table S9.2**: List of the IR active phonon modes of Tu symmetry.

Besides this qualitative difference of the phonon-mode distortion on the molecular levels, we also examined the strength of the induced splitting. From group-theory, the size of the splitting scales with the square of the distortion. Figure S9.1(a) displays how the splitting develops as a function of the fluence of the incoming THz pulse. Each phonon mode distortion was weighted according to its eigenfrequency and mode effective charge in this plot. For the same strength of the driving electric field, the splitting induced by phonon A produces a more significant separation of the $t_{1u}$ levels compared to phonon B. Due to the square scaling of the splitting with the electric field, this effect is further enhanced at higher field strengths.

The computations were performed with the Vienna ab-initio simulation package VASP.6.2[15-17]. For the phonon calculations, we used the Phonopy software package[18] and the Wannier90 package for wannierization[14]. The computations further utilized pseudopotentials generated within the Projected Augmented Wave (PAW) method[18]. Specifically, the following default potentials were used: C $2s^22p^2$ and K $3s^23p^64s^1$. The Generalized Gradient Approximation (GGA[19]) approximation for the exchange-correlation potential was used. For the final numerical setting, a 4x4x4 Monkhorst[20] generated k-point-mesh sampling of the Brillouin zone and a plane-wave energy cutoff of 600 eV were chosen. The calculations were re-iterated self-consistently until the change in total energy converged within $10^{-8}$ eV.

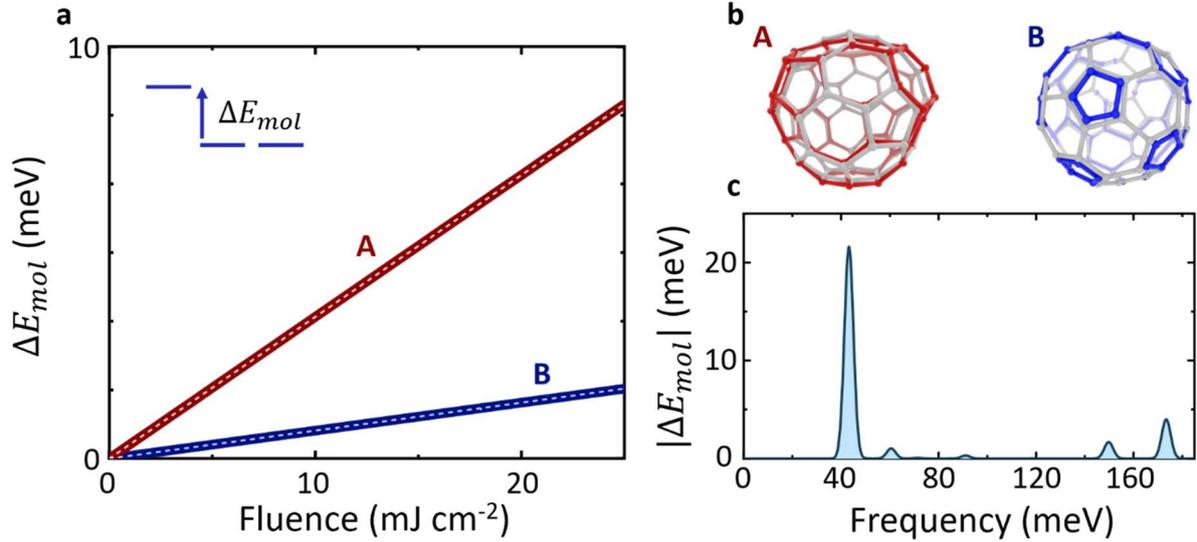

**Figure S9.1: Effect of vibrational distortions on the $t_{1u}$ molecular levels from first-principle computations.** (a) shows the induced splitting of the molecular orbital of $t_{1u}$ symmetry at the Fermi energy (as illustrated by the inset) as a function of drive fluence. The two curves represent the effect of the two distinct $T_{1u}$ IR-phonon modes with eigenfrequencies of 43.2 (red) and 173.4 (blue) meV. The eigen displacement of these modes are shown in (b). Note, that due to the symmetry character of the phonon modes the $t_{1u}$ level split into a single and double degenerate orbital. Lastly, in (c) we show the induced splitting as a function of frequency for a fixed fluence. Here we consider the whole spectrum of $T_{1u}$ IR modes of K3C60, as listed in table S9.2.

## S10. Local electronic hamiltonian calculations

The Hamiltonian proposed in Ref. 21 in order to model superconductivity in alkali-doped fullerides is based on an effective negative Hund's coupling J. It arises from a combination of the usual Hund's coupling with a dynamical Jahn-Teller distortion. This causes states featuring intra-orbital pairing on a buckyball to be energetically favourable. Using ab-initio calculations, values of the intra-orbital interaction U = 0.826 eV and of J = −18.5meV were predicted for $K_3C_{60}$[22]. The phase diagram for the $A_3C_{60}$ family of compounds was computed using DMFT starting from this Hamiltonian and was found to be in quantitative agreement with experimental data[23].

The Hamiltonian can be written as:

$$H = H_{\text{Intra}} + H_{\text{Inter}} + H_{\text{Pairhop}} + H_{\text{Spinswap}}$$

with an intra-orbital interaction with magnitude $U$ given by:

$$H_{\text{Intra}} = U \sum_i^3 n_{i,\uparrow} n_{i,\downarrow}$$

where $n_{i,\sigma} = a_{i,\sigma}^\dagger a_{i,\sigma}$ is the number operator for a spin down electron on orbital $i$ with spin $\sigma \in \{\uparrow,\downarrow\}$. $a_{i,\sigma}^\dagger$ and $a_{i,\sigma}$ are fermion creation and annihilation operators, respectively. The inter-orbital interaction appears as:

$$H_{\text{Inter}} = (U - 2J) \sum_i^3 \sum_j^3 (1 - \delta_{ij}) n_{i,\uparrow} n_{j,\downarrow} + (U - 3J) \sum_\sigma \sum_i^3 \sum_j^{i-1} n_{i,\sigma} n_{j,\sigma}$$

with $\delta_{ij}$ denoting the Kronecker delta. which, given that J is negative, makes these terms higher in energy. In addition, there is a pair hopping term, which corresponds to a transfer of a pair of electrons from one orbital to another. It is given by:

$$H_{\text{Pairhop}} = J \sum_i^3 \sum_j^3 (1 - \delta_{ij}) a_{i,\uparrow}^\dagger a_{i,\downarrow}^\dagger a_{j,\downarrow} a_{j,\uparrow}$$

This term was found to be crucial for the appearance of superconductivity[23]. Finally, there is a "spin swapping" term, where two opposite spins exchange orbitals:

$$-J \sum_i^3 \sum_j^3 (1 - \delta_{ij}) a_{i,\uparrow}^\dagger a_{i,\downarrow} a_{j,\downarrow}^\dagger a_{j,\uparrow}$$

When restricting ourselves to a Hilbert space where the three degenerate orbitals are populated by three electrons (as is appropriate for $A_3C_{60}$ in the atomic limit), we can use a basis given by the different possible arrangements in which the orbitals can be populated:

$\{|\uparrow, \uparrow\downarrow, 0\rangle, |\uparrow, 0, \uparrow\downarrow\rangle, |\uparrow\downarrow, \uparrow, 0\rangle, |0, \uparrow, \uparrow\downarrow\rangle, |\uparrow\downarrow, 0, \uparrow\rangle, |0, \uparrow\downarrow, \uparrow\rangle, |\downarrow, \uparrow, \uparrow\rangle, |\uparrow, \downarrow, \uparrow\rangle, |\uparrow, \uparrow, \downarrow\rangle, |\uparrow, \uparrow, \uparrow\rangle\}$

as well as a second set of states created by flipping all spins in the set above.

In this basis, the Hamiltonian takes on the form:

$$\hat{H} - (3U + 5J)\hat{I} = -J \begin{pmatrix} 0 & -1 & 0 & 0 & 0 & 0 & 0 & 0 & 0 & 0 \\ -1 & 0 & 0 & 0 & 0 & 0 & 0 & 0 & 0 & 0 \\ 0 & 0 & 0 & +1 & 0 & 0 & 0 & 0 & 0 & 0 \\ 0 & 0 & +1 & 0 & 0 & 0 & 0 & 0 & 0 & 0 \\ 0 & 0 & 0 & 0 & 0 & -1 & 0 & 0 & 0 & 0 \\ 0 & 0 & 0 & 0 & -1 & 0 & 0 & 0 & 0 & 0 \\ 0 & 0 & 0 & 0 & 0 & 0 & +2 & -1 & +1 & 0 \\ 0 & 0 & 0 & 0 & 0 & 0 & -1 & +2 & -1 & 0 \\ 0 & 0 & 0 & 0 & 0 & 0 & +1 & -1 & +2 & 0 \\ 0 & 0 & 0 & 0 & 0 & 0 & 0 & 0 & 0 & +4 \end{pmatrix}$$

where $\hat{I}$ is the identity matrix, which encodes an overall energy offset. This matrix is block-diagonal, meaning that different sectors of the Hilbert space are not coupled to each other: For example, there is no term that destroys or creates pairs. Because of the inverted Hund's coupling, i.e. because J is negative, the stretched state $|\uparrow,\uparrow,\uparrow\rangle$ as well as its global spin-flip partner $|\downarrow,\downarrow,\downarrow\rangle$ are now the *most energetic* local eigenstates.

The local ground state is 6-fold degenerate, with an exemplary instance given by: $|g_1\rangle = (|\uparrow,\uparrow\downarrow,0\rangle + |\uparrow,0,\uparrow\downarrow\rangle)/\sqrt{2}$, i.e. it is a state where one singlet pair of electrons has de-localized over two orbitals.

The first excited manifold is 10-fold degenerate. Six of those states are of the type $|e_1\rangle = ((|\uparrow,\uparrow\downarrow,0\rangle - |\uparrow,0,\uparrow\downarrow\rangle)/\sqrt{2}$ i.e. identical to the ground state except for the phase of the de-localized singlet pair (and hence corresponding to a different local angular momentum) – as illustrated in Figure S10.1.

The energy difference between these two manifolds is given by 2*J*=37meV, remarkably close to the observed resonance in the experiment. However, several questions remain in order to determine whether an excitation of this transition is responsible for the experimental observation:

Firstly, how does the light field of the laser couple to this excitation? As the size of a buckyball is comparable to the distance between buckyballs both inter-site and intra-site driving terms may be comparable in terms of the associated energy. Understanding possible inter-site driving terms (arising from the oscillating energy difference between neighbouring sites, given by the electric field multiplied with the charge and the lattice spacing) will require a calculation featuring multiple buckyballs. Locally, because the dynamical Jahn-Teller distortion causes the populated orbitals to be superpositions of several undistorted orbitals, we may expect the electric field to lift the orbital degeneracy, for example through an orbital offset term of the type $H_{\text{offset}} = \Delta(n_{3,\uparrow} + n_{3,\downarrow})$, where $\Delta$ encodes the amplitude of the drive and is oscillating in time. Such a term would in fact cause an excitation from $|e_1\rangle$ to $|g_1\rangle$, but it would not populate any un-paired state (which are not affected by this driving term, as all orbitals are equally occupied).

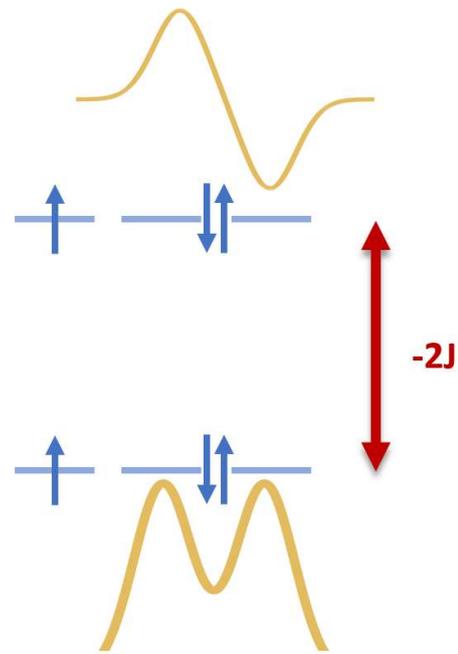

**Figure S10.1: Ground state and first excited state of the local Hamiltonian.** The yellow lines indicate the phase of the pair which is de-localized over two orbitals. The energy spacing between these two states is given by -2$J$

Secondly, $K_3C_{60}$ has an electronic bandwidth of about 0.5eV[23], meaning that the system is far away from the atomic limit (i.e. zero inter-site tunneling). Nevertheless, because the excitation here does not require inter-site tunneling (unlike e.g., double occupancy creation in a regular one-band Hubbard model), it may remain sufficiently separable.

Finally, how does this excitation generate superconductivity? Indeed, the Suhl-Kondo mechanism suggests that in a multi-band system, pairs in any local superposition can contribute to superconductivity, but how the generation of excited-state pairs can lead to superconducting properties starting from a normal state remains to be investigated.